# An On-Demand Optical Quantum Random Number Generator with In-Future Action and Ultra-Fast Response

Mario Stipčević[1,]*, Rupert Ursin[2]

[1]Photonics and Quantum Optics Unit, Center of Excellence for Advanced Materials and Sensing Devices, Ruđer Bošković Institute, Bijenička 54, 10000 Zagreb, Croatia

[2]Institute for Quantum Optics and Quantum Information, Austrian Academy of Sciences, Boltzmanngasse 3, 1090 Vienna, Austria

*E-mail: stipcevi@irb.hr

*Abstract.* **Random numbers are essential for our modern information based society e.g. in cryptography. Unlike frequently used pseudo-random generators, physical random number generators do not depend on complex algorithms but rather on a physicsal process to provide true randomness. Quantum random number generators (QRNG) do rely on a process, which can be described by a probabilistic theory only, even in principle. Here we present a conceptually simple implementation, which offers a 100% efficiency of producing a random bit upon a request and simultaneously exhibits an ultra low latency. A careful technical and statistical analysis demonstrates its robustness against imperfections of the actual implemented technology and enables to quickly estimate randomness of very long sequences. Generated random numbers pass standard statistical tests without any post-processing. The setup described, as well as the theory presented here, demonstrate the maturity and overall understanding of the technology.**

**Introduction**

Digital data processing in computers, mobile devices, ATM machines etc., do have a huge impact on our information-based society. Random numbers are essential for cryptographic protocols which are necessary to ensure security, privacy and integrity of communicated data. In contrast to computational methods used by pseudo-random number generators, physical random numbers generators derive random numbers from a physical source of reasonably random process e.g. flipping a coin. However, systems relying on classical motion actually do have a component of deterministic prediction that will be transferred to the random numbers obtained thereof. On the other extreme is the quantum theory, a branch of physics that strives to understand and predict the properties and behavior of tiny objects, such as elementary particles. One intriguing aspect of the theory is that properties of a particle are not determined with arbitrary precision until one measures them, consequently the individual result of a measurement remains random. This characteristic of the theory describing certain processes provides fundamental randomness that can be used for generating random numbers which are an essential resource for many important applications such as: cryptography, online gambling, Monte Carlo modeling of natural phenomena, randomized algorithms and scientific research. We present a novel type of QRNG whose randomness can be obtained by suitable tuning the device controllable parameters in function of the hardware imperfections. It is unique in simultaneously satisfying three characteristics: (1) a very short latency between the random bit request signal and moment when the bit is generated of (9.8 ± 0.2) ns; (2) all physical processes relevant to generation of a bit happen after the request signal;

(3) a 100% efficiency of producing a bit upon a request. This makes it suitable even for most demanding applications such as loophole-free Bell test. On top of that, we estimate deviation of the QRNG from perfect randomness and demonstrate that generated sequences of random bits pass NIST Statistical Test Suite (STS) [9] without post-processing.

Physical RNGs can be divided into two broad categories: firstly *continuous* which produce random numbers at their own pace and secondly *triggered* which produce a random number upon a request after a bounded time (latency). Both, continuous and triggered RNGs feature the Strobe output which generates a short logic pulse when the new random bit is available at the Random Bit output. Additionally, the triggered type features a Request input. When a pulse is sent to that input it triggers a series of physical events and measurements - resulting in generation of a new random bit. Examples of continuous generators include those that extract random numbers from time-wise random events such as radioactive decay [1], photon arrival [2], or beamsplitter based [3-4] RNG's. Examples of a triggered RNG include sampled time-wise random toggling flip-flop [5-6]. An important consideration is the latency between a moment of request and the moment when the random bit is available for readout (technically the delay between the Request and the Strobe pulses).

An interesting further requirement does come from experimental loophole-free Bell inequality tests. Bell test allows distinguishing quantum mechanics from local hidden variable theories. These experiments are also quite important for future implementation of quantum key distribution devices [7]. Experimental tests performed so far do suffer from so called "loopholes" [8]. In order to close the "locality" [9] as well as the "freedom-of-choice" loophole [10] one needs to decide on random setting of detection basis by means of a RNG that satisfies three properties: (1) all physical processes required for production of a bit must happen completely *in the future* of the trigger, that is anything that happened before the trigger must not have any influence on the generated bit value; (2) a random bit is produced upon a request with certainty within a *bounded* time; (3) in order to facilitate realistic experimental implementation of a loophole-free Bell test, including detection loophole [11], [12] the delay should be less than a few tens of nanoseconds. None of the generators or generating principles known so far satisfies all those requirements simultaneously to that extent.

We present a novel quantum random number generator that has all three above mentioned characteristics guaranteed by design simultaneously. Shown in Fig. 1, it comprises a bit request input (Trigger Input), a laser diode (LD), a single photon detector (PD), and a coincidence circuit consisting of a single AND gate. It functions in the following way. The external trigger signal causes LD to emit a short (sub-nanosecond) light pulse. We define that one random bit is generated upon *every* trigger signal. The value of the random bit is defined as the state of the detector's output at the moment of positive-going edge of the synchronous Strobe signal which is derived from the Trigger signal by a suitable delay (latency). Note, if emission and detection of light were classical processes then detection would either happen every time (if pulse energy is higher than some given threshold) or never (if below the threshold). However, due to the quantum nature of light, detection of a photon arising from the laser pulse is a binomial process with success probability $p_1$ that can take on any value in the range [0, 1]. The energy of the light pulse falling upon the detector is carefully set such that the probability $p_1$ of detecting a photon (and thus generating a value of "1") is as close as possible to $p_1 = 0.5$. We assumed that the laser is stable in power and the detectors efficiency is constant during the measurement time. Note, the detection efficiency of the chosen PD is irrelevant since it is always possible to set pulse power such that the above condition is met. This is in contrast with e.g. pulsed beam-splitter method [4] where efficiency of detector affects the bit generation rate. For each and every trigger signal, we get an answer from the QRNG, hence we call the device 100% efficient.

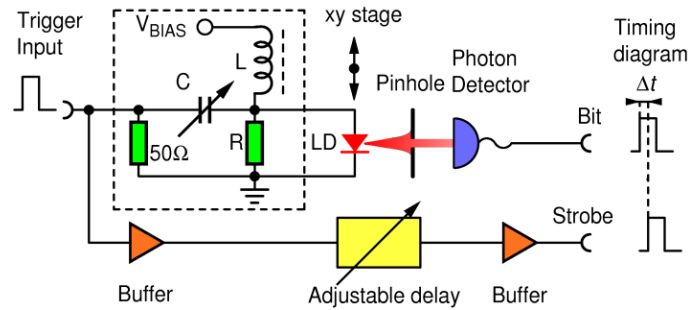

Fig. 1. Experimental realization of the triggered short-latency in-future-action quantum random number generator. A trigger at the input is generating the Strobe signal and in parallel triggers a laser pulse from the laser diode LD powering it via a circuit consisting of the resistor R, the inductor L, the variable capacitor C and bias voltage $V_{BIAS}$. The laser diode is mounted on an XY direction translation stage and can move relative to a 50 µm pinhole placed in front of the photon detector thus allowing for fine adjustment of the optical coupling to the photon detector. The delay is essentially determined by the propagation time of the photon detector while all processes relevant to the bit value happen in future of the Trigger signal.

Under the assumption that both, the light source and the detector are completely reset to their initial conditions between subsequent triggers it is impossible for generated bit values to "communicate", i.e. influence each other. Consequently there would be no correlation among successive bits. Having these two characteristics (probability of ones equal to 0.5 and absence of correlation among successive bits) a pool of generated bits has no other possibility than to be random [31]. It is assumed that a bit generated upon a trigger has no history prior to that trigger since all relevant physical processes, namely: (1) powering of the laser diode and subsequent light pulse emission, (2) photon detection and (3) detector-trigger coincidence, are all happening *after* the trigger. The efficiency of the method is high: two random bits per photon detection as compared to $\leq 1$ bit for beamsplitter [4] and $\leq 0.5$ for arrival-time [2] methods. Even though it does not allow for higher bit generation rate because the ultimate rate is bounded by inverse of the dead time, it does put a less strain to the detector reducing its power consumption and possibly extending its lifetime.

**Results**

In the experimental realization of the RNG, shown in Fig. 1, light pulses are obtained from a single mode laser diode LD (Sony DL3148-025 at 650 nm) driven by a sub-nanosecond current pulse formed by a simple RLC circuit upon each positive-going edge of the trigger pulse. Passive driver design ensures smallest delay between the driving electrical pulse and the light pulse. Coarse adjustment of the energy and width of light pulses is made by the variable capacitor C. The laser diode is mounted on an XY translation stage and can move relative to a 50 µm pinhole placed in front of the photon detector thus allowing for fine adjustment of the optical coupling and in turn the detection probability $p_1$.

The laser pulse features a jitter of 190 ps FWHM with respect to the trigger raising-edge. In order to avoid degradation of pulse power and shape, shortest period between two consecutive triggers should be ≥ 40 ns. The photon detector is home-made and makes use of a SLiK silicon avalanche photodiode (APD) recovered from a PerkinElmer SPCM-AQR module complemented by an active avalanche quenching circuit (AQC) described in Ref. [13]. For lower dark counts and stable performance the APD is cooled to -10°C. Characteristics of the detector are: output pulse width $\tau_{pd}$ = 8 ns, dark counts of 235 cps, dead time of $\tau_{dead}$ = 22 ns, detection efficiency of 65% at 650 nm and jitter of about 320 ps FWHM. The distinctive characteristic of this AQC is that the delay between photon detection and the output pulse of the PD is only about 5 ns. Total delay between the trigger input and output of PD is measured to be (6.5 ± 0.2) ns with a jitter of (370 ± 50) ps FWHM. Because of this jitter, the Strobe signal should appear at least $\Delta t$ = 2 ns later than the detector's output to ensure high efficiency of picking up the detection signal. Therefore, the latency between the Trigger input and Strobe output was fixed to 8.5 ns by means of the adjustable electrical delay shown in Fig. 1.

While in theory there should be no correlation among the bits, due to inevitable memory effects in realistic devices some autocorrelation appears also in experimental realization of the QRNG. Successive pulses of a pulsed laser diode are phase randomized exhibiting a Poisson statistics of number of emitted photons per pulse ($n$) [23-24]. The detection of such a state is ether supposed to be ballistic ($n$ independent detection trials) or superlinear [14]. Crucial insight into the present QRNG is that any details of photon emission or detection are irrelevant as long as all physical processes pertaining to one emission and subsequent detection event are completed (i.e. die off) before the next trigger. This would ensure no correlations among generated bits. However, while the turn-on and turn-off processes in a laser diode have typical lifetimes on the order of <100 ps [15], a photon detection imperfections (dark counts, dead time, afterpulsing) involve effects on a time scale of tens to hundreds of nanoseconds that ultimately limit the achievable trigger rate and randomness. Dark counts are randomly distributed in time and therefore do not carry *per se* any correlating information and are furthermore greatly suppressed by tight coincidence between trigger and detector pulses. However, dead time and afterpulsing may cause correlations among bits. Since afterpulsing probability of the used APD dies-off nearly exponentially in time [16], in the limit of long enough trigger period, only neighboring bits may be non-negligibly correlated. Under that condition, correlations among bits is characterized by the serial autocorrelation between neighboring bits, that is coefficient $a_1$ defined as [17]:

$$a_k = \frac{\sum_{i=1}^{N-k}(x_i - \bar{x})(x_{i+k} - \bar{x})}{\sum_{i=1}^{N-k}(x_i - \bar{x})^2} \qquad (1)$$

where $x_i$ are generated bits and lag $k = 1$. Throughout the paper we use statistics of $N = 10^9$ bits for each measurement point, leading to statistical error of $1/\sqrt{N-k} \approx 3.2 \cdot 10^{-5}$. Random bits have been generated upon a periodic trigger with frequency spanning from 1 to 25 MHz. Statistical bias defined as $b = p_1 - 0.5$, was manually adjusted to zero within ±0.0005 before each measurement point. The generated bits were transferred to a PC computer via a USB2 controller. Correlation coefficient $a_1$ has been evaluated using ENT software [18]. Results are shown as hollow dots in Fig. 2.

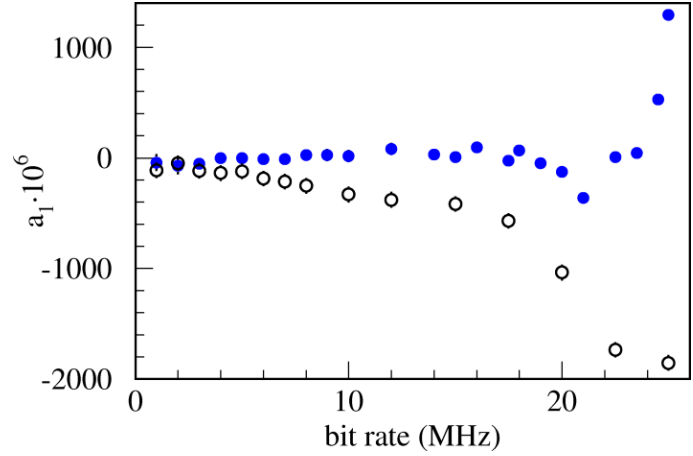

Fig. 2. A series of autocorrelation coefficients $a_1$ as a function the triggered bit rate, measured for two distinct detector pulse widths ($\tau_{pd}$): 8 ns (hollow dots) and 21 ns (filled dots). Statistics per coefficient is $10^9$ bits. One sigma error bars are barely visible being roughly equal to the dot size.

We see that $a_1$ is generally small, negative and that its magnitude rises with the rate. To explain this behavior we start by considering a successful detection of a photon (bit value "1") as shown in Fig. 3.

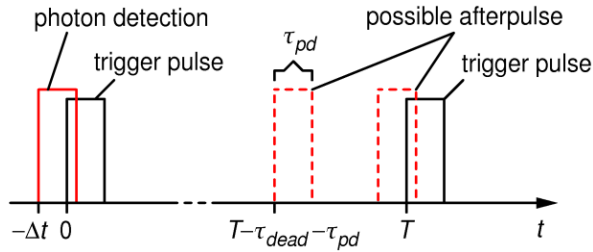

Fig. 3. Time sequence of a detection and possible locations of an afterpulse event that would cause correlation between subsequent random bits.

The next bit value is requested/generated a period *T* later. Afterpulsing in conjecture with dead time causes two competing effects. First, at time *T* there will be an enhanced probability $P_+$ to generate "1" due to an afterpulse appearing in coincidence with the trigger. Second, with probability $P_-$, an afterpulse appearing less that the dead time $\tau_{dead}$ and *before* the trigger will cause the detector to miss the next photon whose probability would otherwise be ½. The total correlation is then given as:

$$a_1 = \frac{1}{2}[P_+ - P_-] = \frac{1}{2}\left[\int_{T+\Delta t-\tau_{pd}}^{T+\Delta t} P_a(t)dt - \frac{1}{2}\int_{T+\Delta t-\tau_{pd}-\tau_{dead}}^{T+\Delta t-\tau_{pd}} P_a(t)dt\right] \quad (2)$$

where $P_a(t)$ is probability density function for appearance of an afterpulse at time *t* after a detection event. The overall factor ½ stems from the fact that two photons are generated on average per photon detection. In our case, where $\tau_{pd}$ = 8 ns and $\tau_{dead}$ = 22 ns, the net autocorrelation $a_1$ is negative because the integration interval of the second term (of length $\tau_{dead}$) is longer than that of the first term (length $\tau_{pd}$) and because $P_a(t)$ is larger in the second integral. However, since the two integrals are the contiguous parts of an integral over a fixed interval (of length $\tau_{pd} + \tau_{dead}$) it could be possible to choose $\tau_{pd}$ such that the correlation vanishes. If a simple exponential model of afterpulsing is assumed, i.e. $P_a(t) = \frac{P}{\tau_a} e^{-t/\tau_a}$ [19] where $P$ is the total afterpulsing probability, by requiring $a_1 = 0$ one gets:

$$e^{\frac{\tau_{pd}}{\tau_a}} \left[ 3 - e^{\frac{\tau_{dead}}{\tau_a}} \right] = 2 \qquad (3)$$

from which $\tau_{pd}$ can be expressed as:

$$\tau_{pd} = \tau_a \ln \left[ \frac{2}{3 - e^{\frac{\tau_{dead}}{\tau_a}}} \right]. \qquad (4)$$

Interestingly, for a hypothetical detector with a vanishing afterpulsing probability (i.e. $\tau_a \to \infty$) Eq. (3) would be automatically satisfied and any value of $\tau_{pd}$ would be optimal. For our particular SLiK diode we measured $\tau_a = 33$ ns and $P = 0.047$. Inserting $\tau_a$ and $\tau_{dead}$ in Eq. (4) yields $\tau_{pd} \approx 21$ ns. Apparently, the value of $\tau_{pd}$ optimal for cancelation of $a_1$ is independent of $T$. To verify that experimentally we vary the width of the detector's output pulse at the AQC and a evaluate autocorrelation as a function of $\tau_{pd}$ for several bit rates (10 MHz, 15 MHz, 17.5 MHz and 20 MHz). Experimental results shown in Fig. 4 indicate that an overall minimum of the autocorrelation is indeed obtained for $\tau_{pd} \approx 21$ ns and that is rather insensitive on the bit rate.

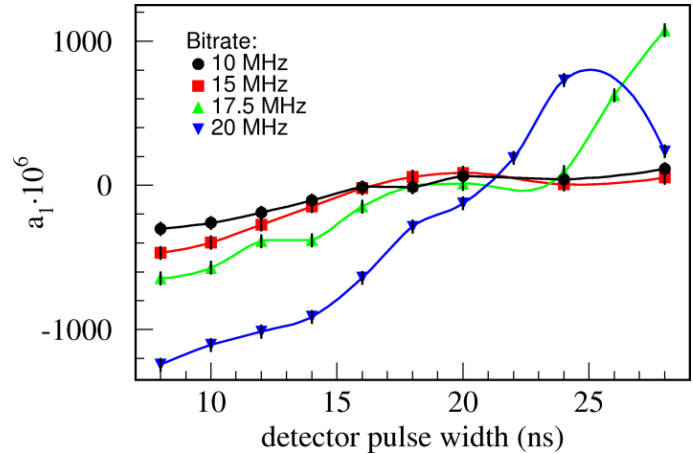

Fig. 4. Autocorrelation coefficients $a_1$ as a function of detector's pulse width ($\tau_{pd}$), measured for a set of bit rates. An overall minimum is obtained for $\tau_{pd} \approx 21$ ns. Statistics per measurement is $10^9$ bits. One sigma error bars are barely visible being roughly equal to the dot size.

We further note that following a detection of a photon at $-\Delta t$, the detector goes into the dead time and therefore afterpulses would contribute to the second integral in Eq. (1) only if its starting range $(T + \Delta t - \tau_{pd} - \tau_{dead})$ is greater than $\tau_{dead}$, that is:

$$T > 2\tau_{dead} + \tau_{pd} - \Delta t \qquad (5)$$

which corresponds to bit rate of about $1/T < 16$ MHz. For higher trigger rates the second integral in Eq. (1) would become smaller and the autocorrelation would rise sharply, as indeed observed for bitrates of 17.5 MHz and 20 MHz. We note that higher lag coefficients ($k > 1$) are obtained by shifting the boundaries of both integrals in Eq. 2 by $T$, that is: $a_k = a_1 \exp\left(\frac{(k-1)T}{\tau_a}\right)$.

After setting $\tau_{pd}$ to the optimal value of 21 ns, correlation coefficient $a_1$ has been evaluated again as a function of bit generation rates in the range 1-25 MHz. Results displayed in Fig. 2 (dots) show a significant improvement with respect to the result obtained with the original pulse width of 8 ns (circles). The absolute value of $a_1$ is less than $1.25 \cdot 10^{-4}$ for bit rates all the way up to 20 MHz. At higher rates correlation quickly diverges because our simple model fails due to the effects explained above and possibly other smaller imperfections not taken into account.

In practice Eq. (4) cannot be exactly satisfied for physical devices. It is therefore interesting to investigate the sensitivity of autocorrelation to variation of parameters such as detector pulse width ($\tau_{pd}$), dead time ($\tau_{dead}$) and bit generation period ($T$). By substituting the exponential afterpulsing model in Eq. (2) and taking partial derivative of $a_1$ with respect to $\tau_{pd}$ we get:

$$\frac{\partial a_1}{\partial \tau_{pd}} = \frac{P}{4\tau_{pd}}\left[3 - e^{\tau_{dead}/\tau_a}\right]e^{-(T+\Delta t - \tau_{pd})/\tau_a}. \tag{6}$$

Evaluated at $\tau_{pd} = 21$ ns, for $T = 100$ ns, $\tau_{dead} = 22$ ns, $\tau_a = 33$ ns, $\Delta t = 2$ ns and $P = 0.047$, Eq. (6) predicts sensitivity of $a_1$ with respect to $\tau_{pd}$ of $32 \cdot 10^{-6}$ ns$^{-1}$ which is indeed in a good agreement with the slope of the 10 MHz curve in Fig. 4. Similar analysis for dead time yields a sensitivity of $-59 \cdot 10^{-6}$ ns$^{-1}$, whereas for generation period the variation sensitivity is $0.2 \cdot 10^{-6}$ ns$^{-1}$ only. Since the three parameters ($\tau_{pd}$, $\tau_{dead}$, $T$) can be engineered with high precision and stability on the order of 1 ns, randomness quality of the present generator is predominantly affected by stability of bias which is about $500 \cdot 10^{-6}$. We find that serial correlation coefficients $a_k$ with lag $1 < k \leq 64$ are consistent with zero within statistical error for $T = 100$ ns and $N = 10^9$. This is to be expected since with every lag the afterpulsing probability (and consequently the serial correlation) drops roughly by a factor of $\exp(T/\tau_a) \approx 21$, so that the second and all further serial coefficients are much smaller than our statistical error.

In order to further improve on both the statistical bias and the autocorrelation, one could use the Von Neumann extractor [23]. However, while on average it takes a block of 4 bits to generate one bit, the required block length can span anywhere from 2 bits to infinity before the next output bit is generated. In our case that would result in lowering of the bit production efficiency to only 25% and enlargement of the delay between the request and availability of the random bit. Therefore we chose an alternative, well known approach, which enabled us to keep the 100% efficiency and bounded latency: we built two independent generators of the type shown in Fig. 1, distributed the same trigger signal to their inputs and logically XORed their outputs. The XOR gate added another 1.3 ns of propagation delay, therefore the delay between the Trigger and Strobe was enlarged by the same amount, i.e. to 9.8 ns. According to

[20] XORing two independent random strings each with bias $b$ and autocorrelation $a_1$ results in a new string with an improved bias $b'$ and autocorrelation $a_1'$ :

$$b' = -2b^2 \qquad (7)$$

$$a_1' = a_1^2 + 8a_1 b^2 \qquad (8)$$

At 10 Mbit/s (i.e. *T* = 100 ns) for a single QRNG we measured: $b \leq 5 \cdot 10^{-4}$; $a_1 \leq 5 \cdot 10^{-5}$. Higher lag correlations were consistent with zero, within statistical errors, as expected in our model. By applying Eqs. (7-8) we estimate the upper bounds for the residual bias and autocorrelation of the XORed QRNGs to be: $|b'| \leq 5 \cdot 10^{-7}$ and $|a_1'| \leq 3 \cdot 10^{-9}$, respectively.

In our model, explained in Fig. 3, there are no deviations from randomness other than bias and serial autocorrelation and we saw that coefficients with lag $k > 2$ contribute negligibly both theoretically and as confirmed by measurements. To detect statistically the above imperfections as a 3 sigma effect, one would need to generate at least $10^{13}$ bits for bias, and $10^{18}$ for correlation, showing that bias is the leading imperfection. However, afterpulsing is generally more complex [19] and there could be other small imperfections in the setup that were not accounted for in our model, all of which could limit the achievable randomness.

In order to demonstrate that our random bits pass traditional statistical tests, several sequences of $10^9$ bits (1000 samples of 1 Mbits) were generated by the XORed QRNG at 10 Mbit/s and verified to indeed pass the NIST STS with high scores. We used plain data directly coming out of the device without any post-processing. Typical results are shown in Table 1.

| Statistical test | *p*-value | Proportion/Threshold | Result |
|---|---|---|---|
| Frequency | 0.784927 | 994/980 | Pass |
| Block frequency | 0.096578 | 992/980 | Pass |
| Cumulative sums | 0.767582 | 997/980 | Pass |
| Runs | 0.775337 | 995/980 | Pass |
| LongestRun | 0.103138 | 991/980 | Pass |
| Rank | 0.657933 | 994/980 | Pass |
| FFT | 0.251837 | 993/980 | Pass |
| NonOverlappingTemplate | 0.574903 | 994/980 | Pass |
| OverlappingTemplate | 0.867692 | 987/980 | Pass |
| Universal | 0.697257 | 994/980 | Pass |
| ApproximateEntropy | 0.348869 | 993/980 | Pass |
| RandomExcursions | 0.588541 | 626/615 | Pass |
| RandomExcursionsVariant | 0.235040 | 625/615 | Pass |
| Serial | 0.637119 | 990/980 | Pass |
| LinearComplexity | 0.880145 | 986/980 | Pass |

Table 1. Typical results of NIST statistical test suite STS-2.1.1 for 1000 samples of 1 Mbits generated with the XORed QRNG. For each statistical test an overall *p*-value as well as proportion of samples that passed the test versus theoretical threshold are given.

Finally, as an alternative approach to improve randomness, non-overlapping pairs of bits from a single QRNG operated at 10 Mbit/s have been XORed. In that case, the resulting bias and correlation are given by [20]:

$$b' \approx -2b^2 - a_1/2 \tag{9}$$

$$a_1' \approx 4a_1 b^2 \tag{10}$$

which gives $b' \approx -2.6 \cdot 10^{-6}$ and $a_1' \approx 5 \cdot 10^{-11}$. Again, 1000 samples of 1 Mbits have passed NIST test suite. The drawback of this approach is halving of the effective bit rate (to 5 Mbits/s) and doubling the latency, while the good side is requirement for only one photon detector.

**Discussion**

A conceptually simple, on-demand optical quantum random number generator is presented that simultaneously features: (1) ultra-fast response upon a bit request (9.8 ns), (2) 100% bit generation efficiency upon the trigger and (3) in-future-of-request random action. While its characteristics are of particular relevance to some applications (such as Bell tests or random logic [25]), it can be used for a much wider range of applications. It can deliver random bits at a maximum rate of currently 10 MHz featuring very low randomness errors without post-processing. Sources of randomness errors and their sensitivity to variations in hardware components have been studied, modeled and shown to be small. In comparison, other post-processing free-running QRNGs have achieved 100% efficiency and nanosecond scale response by quick sampling of a randomly toggling flip-flop [6], [26], but with all relevant physical processes happening hundreds of nanoseconds in the past of the request due to long delays in optical and electrical paths or long range correlations among bits. A post-processing-free QRNG based on self-differencing technique [27] operated at a clock 1.03 GHz delivers bits randomly at an average rate of 4.01 Mbit/s thus having efficiency of only about 4‰. In a setup having a similar topology to ours [28] a gain-switched laser diode feeds an asymmetric Mach-Zender interferometer whose output intensity is measured by a photodiode and digitized by 8-bit ADC, whereas in [29] an in-future-of-request continuous-variable QRNG is based on phase diffusion in a laser diode. Both QRNGs feature unavoidable requirement for ADC conversion followed by complex post-processing which results in long response times. Furthermore, none of the above discussed constructs has been tested random for strings longer than $\sim 10^9$ bits, which can be too short for applications like Monte Carlo calculations and simulations. For the XORed QRNG, assuming the validity of our model, we estimated that randomness imperfections can not be statistically detected for a sequence of generated bits shorter than $\sim 10^{13}$ bits. A notable success in randomness estimation is achieved in [30] by calculating propagation of min-entropy through privacy amplification claiming randomness for strings of up to $\sim 10^{96}$ bits, but at the expense of time-consuming post-processing and long history of physical events prior to the bit request. Finally, achieved delay between a request and availability of random bit in our QRNG is arguably the shortest possible with a given state of technology since only a logically minimal sequence of processes is required to generate one bit, namely a light pulse emission followed by a photon detection. The presented bit generating method in principle allows for miniaturization of the QRNG to a chip level with the existing technology. This opens possibility for wider range of applications.

**Methods**

All logic circuits required for the RNG as well as data acquisition are made within a single Altera MAX3000 family reconfigurable chip complemented with a Cypress CY7C68013 communication chip for transfer of data to the PC computer via USB2 link. Statistical analysis of random bits is performed using ENT [18] and NIST Statistical Test Suite version 2.1.1 [9] software.

**References**

1. Figotin A. *et al.*, inventors; The Regents of the University of California, asignee; *A random number generator based on spontaneous alpha-decay*. PCT patent application WO0038037A1.
2. Stipčević M., Medved Rogina B., Quantum random number generator based on photonic emission in semiconductors, *Rev. Sci. Instrum.* **78**, 045104:1-7 (2007).
3. Rarity J. G., Owens P. C. M., Tapster P. R., Quantum random-number generator and key sharing, *J. Mod. Opt.* **41**, 2435-2444 (1994).
4. Stefanov A., Gisin N., Guinnard O., Guinnard L., Zbinden H., Optical quantum random number generator, *J. Mod. Opt.* **47**, 595-598 (2000).
5. Fürst H. *et al.*, High speed optical quantum random number generation, *Opt. Express* **18**, 13029-37 (2010).
6. Stipčević M., Fast nondeterministic random bit generator based on weakly correlated physical events, *Rev. Sci. Instrum.* **75**, 4442-4449 (2004).
7. Scarani V. *et al.*, The security of practical quantum key distribution, *Rev. Mod. Phys.* **81**, 1301–1350 (2009).
8. Merali Z., Quantum mechanics braces for the ultimate test, *Science* **331**, 1380–1382 (2011).
9. Weihs G., Jennewein T., Simon C., Weinfurter H., Zeilinger A., Violation of Bell's Inequality under Strict Einstein Locality Conditions, *Phys. Rev. Lett.* **81**, 5039–5043 (1998).
10. Scheidl, T. *et al.*, "Violation of local realism with freedom of choice," *Proc. National Academy of Sciences* **107**, 19708–19713 (2010).
11. Giustina M. et al., Bell violation using entangled photons without the fair-sampling assumption, *Nature* **497**, 227–239 (2013).
12. Christensen B. G. et al., Detection-Loophole-Free Test of Quantum Nonlocality, and Applications, *Phys. Rev. Lett.* **111**, 130406 (2013)
13. Stipčević M., Active quenching circuit for single-photon detection with Geiger mode avalanche photodiodes, *Appl. Opt.* **48**, 1705-1714 (2009).
14. Lydersen L. *et al.*, Superlinear threshold detectors in quantum cryptography, *Phys. Rev. A* **84**, 032320 (2011).
15. Pesquera L., Revuelta J., Valle A., and Rodriguez M. A., *Theoretical calculation of turn-on delay time statistics of lasers under PRWM*, Proc. SPIE 2994: Physics and Simulation of Optoelectronic Devices V [Osinski M., Chow W. W., (eds.)] (SPIE, San Jose, 1997).
16. Stipčević M., Gauthier D. J., *Precise Monte Carlo Simulation of Single-Photon Detectors*, Proc. SPIE Vol. 8727: Advanced Photon Counting Techniques VII [Itzler M. A., Campbell J. C. (eds.)] (SPIE, Baltimore, 2013).
17. Knuth D., *The art of computer programming Volume 2: Seminumerical Algorithms*, Third Edition [70-71] (Addison-Wesley, Reading, 1997)
18. Walker, J., *ENT - A Pseudorandom Number Sequence Test Program*, (2003) URL: http://www.fourmilab.ch/random/, Date of access: 05/02/2014.
19. Giudice A. C., Ghioni M., and Cova S., *A process and deep level evaluation tool: afterpulsing in avalanche junctions*, Proc. European Solid-State Device Research 2003 (ESSDERC 03). 16–18 Sept. 2003 p. 347–350.
20. Davies R., *Exclusive OR (XOR) and hardware random number generators*, February 28, 2002, URL: http://www.robertnz.net/pdf/xor2.pdf, Date of access: 05/02/2014.
21. Rukhin A. *et al.*, NIST Special Publication 800-22rev1a (April 2010), URL: http://csrc.nist.gov/rng , Date of access: 01/02/2012.
22. Von Neumann J., *Various techniques for use in connection with random digits*, [Von Neumann Collected Works, Vol 5] [768-770] (Macmillan, New York, 1963).
23. Henry C., "Theory of the line width of semiconductor lasers," *IEEE J. Quantum Electron.* **18**, 259–264 (1982).
24. Henry C., "Phase noise in semiconductor lasers," *J. Lightwave Technol.* **4**, 298–311(1986).

25. Stipčević M., "Quantum random flip-flop based on random photon emitter and its applications", arXiv:1308.5719 [quant-ph]
26. Jennewein T., Achleitner U., Weihs G., Weinfurter H., Zeilinger A., "A Fast and Compact Quantum Random Number Generator", *Rev. Sci. Instrum.* **71**, 1675-1680 (2000).
27. Dynes J. F., Yuan Z. L., Sharpe A. W., and Shields A. J., "A high speed, postprocessing free, quantum random number generator", *Appl. Phys. Lett.* **93**, 0311109 (2008).
28. Yuan, Z. L. et al., "Robust random number generation using steady-state emission of gain-switched laser diodes", *Appl. Phys. Lett.* **104**, 261112 (2014).
29. Abellán C. et al., "Ultra-fast quantum randomness generation by accelerated phase diffusion in a pulsed laser diode", *Opt. Express* **22**, 1645-1654 (2014).
30. Sanguinetti B., Martin A., Zbinden H., and Gisin N., "Quantum Random Number Generation on a Mobile Phone", *Phys. Rev. X* **4**, 031056 (2014).
31. D. Frauchiger, R. Renner, and M. Troyer, "True randomness from realistic quantum devices," SPIE Security+Defense Conference Proceedings, Volume 8899 (2013), arXiv:1311.4547 [quant-ph].

**Acknowledgements**

We acknowledge help of the Croatian Ministry of Science, Education and Sports project 098-0352851-2873. We also thank EC project QESSENCE (number 15848) and the FFG project Nr. 4299236.

**Author contributions**

M.S. designed the QRNG and run the experiments. M.S. and R.U. wrote the main manuscript text. Both authors reviewed the manuscript.

**Additional information**

Competing financial interests: The authors declare no competing financial interests.